\documentclass[prl,twocolumn,showpacs,10pt]{revtex4}
\usepackage[dvipdfm]{graphicx}
\usepackage{amsmath,amssymb}
\usepackage{bm}
\usepackage[T1]{fontenc}

\usepackage{graphicx}

\newcommand{\ra}{\rangle}

\newcommand{\eps}{\varepsilon}

\def\hto0{\xrightarrow{h\to 0}}

\newcommand{\defeq}{\stackrel{\rm{def}}{=}}

\newcommand{\bge}{\begin{equation}}
\newcommand{\ede}{\end{equation}}
\newcommand{\be}{\begin{equation}}
\newcommand{\ee}{\end{equation}}

\newcommand{\cP}{{\mathcal P}}

\newcommand{\Mak}{ M_N(a,\kappa)}

\newcommand{\ka}{\kappa}

\newcommand{\heff}{\hbar_{eff}}
\newcommand{\bet}{\bm{\eta}}

\newcommand{\IN}{{\mathbb N}}
\def\t2{{\mathbb T}^2}

\begin{document}
\title{Resonance distribution in open quantum chaotic systems}
\author{S. Nonnenmacher and E. Schenck}
\affiliation{Institut de Physique Th\'eorique,
CEA/DSM/IPhT, CEA-Saclay, 91191 Gif-sur-Yvette, France}

\begin{abstract}
In order to study the resonance spectra of 
chaotic cavities subject to some damping (which can be due to absorption
or partial reflection at the boundaries), we use a model of damped quantum maps.
In the high-frequency limit, the distribution
of (quantum) decay rates is shown to cluster near a ``typical'' value, which is
larger than the classical decay rate of the corresponding damped ray dynamics. The speed of
this clustering may be quite slow, which could explain why it has not 
been detected in previous numerical data. 
\end{abstract}

\maketitle

\medskip

Recent experimental and theoretical studies have focussed on the dynamics of
waves inside quasi 2-dimensional cavities which are ``partially open''; this partial opening may be
due to various physical phenomena. For instance, an acoustic wave evolving in air or in
a metallic slab will lose intensity due to friction and heating. 
In a microwave cavity,
the dissipation mostly occurs at the boundary
through ohmic losses. The light propagating inside a dielectric (micro)cavity 
is partially reflected at the boundary, which can be
described as an ``effective damping'' at the boundary. In all these systems, 
the discrete stationary modes
correspond to complex eigenvalues (or resonances) of the form $k_n=\omega_n-i\Gamma_n/2$,
where $\Gamma_n$ is called the {\it decay rate} of the mode. 

When the shape of the cavity induces a {\it chaotic} ray dynamics (e.g. the ``stadium'' shape), 
the eigenvalues $\{k_n\}$ cannot be computed analytically, but 
methods of ``quantum chaos'' can be applied to predict their statistical distribution in
the high-frequency limit $\omega_n\to\infty$. Statistical studies of resonances started 
in the 1960' with initial applications to nuclear physics \cite{moldauer}. 
New applications emerged
when experiments on mesoscopic quantum dots \cite{qu-dots}, 
microwave cavities \cite{microwave} or optical fibers \cite{mortess}
allowed to construct cavities with prescribed geometries, and study the
dependence of the quantum dynamics with respect to this geometry. A recent
interest for dielectric microcavities comes from the potential applications
to microlasers: choosing the shape of the cavity appropriately allows to produce 
a strongly directional emission \cite{Gmachl}. The first step to understand the
(nonlinear) lasing modes is to 
study the passive (resonant) modes of the cavity.

Various dissipation effects have been taken into account by adding to the self-adjoint
Hamiltonian (representing the dissipationless system) an effective 
imaginary part, which describes the coupling between the internal 
cavity modes and the external channels \cite{Weiden}.
One analytical tool to study chaotic cavities has been to replace the Hamiltonian (and sometimes also the 
effective coupling)
by some sort of random matrix: this has lead to
theoretical distributions, which have been favorably 
compared with numerical or experimental spectra \cite{FS,Kottos}. 

In this Rapid Communication we focus on situations where the coupling is strongly nonperturbative,
and is distributed over a large part of the cavity or of its boundary, so that
the number of coupled channels becomes ``macroscopic'' in the high-frequency (semiclassical) limit.
Using a {\em nonrandom} model of {\em damped quantum maps}, we find that, in this limit, 
the distribution of quantum decay rates becomes asymptotically
peaked on a ``typical'' value $\gamma_{typ}$, which is the ergodic mean
of the local damping rate. This
clustering does not seem to appear if one replaces the unitary part of the quantum map 
by a {\em random} unitary matrix, as
is often done in the quantum chaos literature \cite{KNSch,WeiFyod}.
Such a clustering has been rigorously proved for damped waves on ergodic manifolds \cite{Sj}; we believe it to occur
as well in the various types of partially open quantum systems mentioned above.
Yet, the width of the distribution may decay very slowly in the semiclassical limit ($\lesssim (\ln k)^{-1/2}$), 
which could explain why this semiclassical clustering is hardly visible in numerical computations of chaotic 
dielectric cavities \cite{lebental07,shino-hara07,WM} or damped quantum maps \cite{KNSch}; such a slow
decay indeed occurs within a solvable toy model we briefly describe at the end of this note.

\medskip

\begin{figure}
\includegraphics[width=0.4\columnwidth]{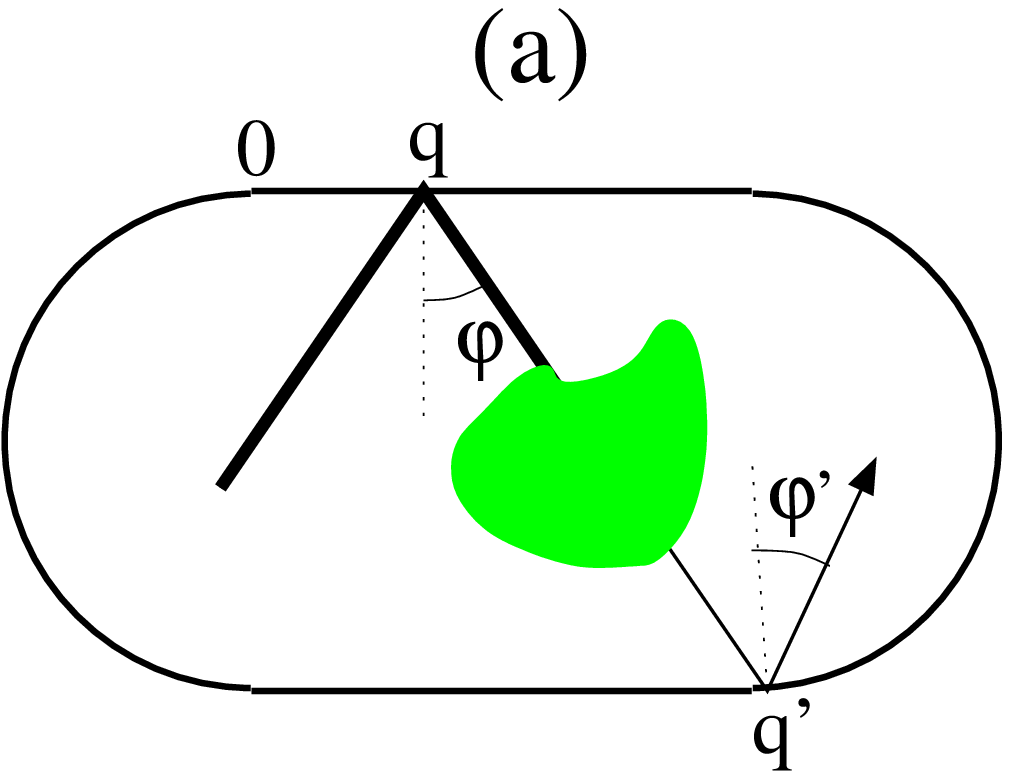}
\includegraphics[width=0.4\columnwidth]{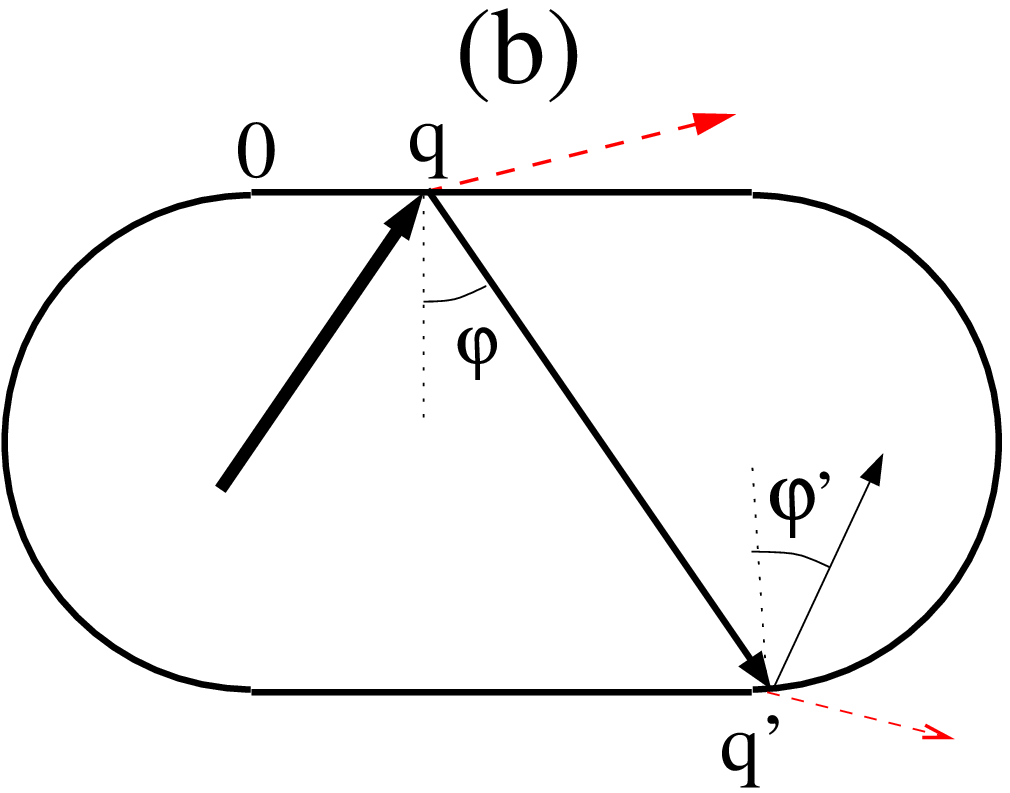}
\caption{(Color online) (a) Weighted ray dynamics inside a cavity 
with inhomogeneous absorption. 
(b) absorption (or partial reflection) at the boundary. The ray intensity corresponds
to its thickness. Dashed lines correspond to refracted rays.
\label{f:damped-rays}}
\end{figure}

Let us now describe the model of damped quantum map, which has been introduced and numerically
investigated in \cite{KNSch} to mimick the resonance spectra of dieletric microcavities.
To motivate this model,  
we first briefly analyse the dynamics of a few cavity wave systems.
The first situation consists in 
a smooth absorption inside the cavity, represented by the 
{\it damped wave equation}
$\big(\partial_t^2-\Delta+2b(x)\partial_t\big)\psi(x,t)=0$.
Here the ``damping function'' $b(x)\geq 0$ measures the local absorption rate. 
A high-frequency wavepacket evolving along a classical trajectory
is continuously damped by a factor $\approx e^{-\int_0^t b(x(s))ds}$. The classical
limit of the dynamics consists in the propagation of rays with decreasing intensity,
also called {\it weighted ray dynamics} (Fig.~\ref{f:damped-rays}, (a)).
When the dissipation occurs
at the boundary (e.g. through ohmic losses), an incident high-frequency wavepacket
hitting the boundary will be reflected,
with its amplitude reduced by a {\em subunitary} factor $a(q,\varphi)$ 
(Fig.~\ref{f:damped-rays}, (b)).
The same phenomenon effectively
occurs in the case of light scattering through a quasi-2D dielectric 
microcavity of optical index $n>1$. 
The rays propagating inside the 
cavity are partially reflected at the boundary,
the remaining part being refracted outside and never return provided the cavity is convex (dashed lines in
Fig.~\ref{f:damped-rays}, (b)).
In the high-frequency limit, the reflection factor is given by 
Fresnel's coefficient, which depends on the light polarization and of $p=\sin\varphi$. For instance, 
in case of transverse magnetic polarization, the coefficient
is a simple complex function $a_{TM}(p)$ \cite{KNSch}, which
has unit modulus when $|p|\geq 1/n$ (full reflection) and is minimal at $a_{TM}(0)=\frac{n-1}{n+1}$.

\medskip

To analyze a 2D classical billiard, it is convenient to reduce the flow
to the {\it bounce map} $\kappa:(q,p=\sin\varphi)\mapsto (q',p'=\sin\varphi')$, 
which acts canonically on the boundary phase space. 
At the quantum level, the spectrum of the
closed cavity can be obtained by studying a $k$-dependent integral operator acting on the 
boundary, which effectively quantizes the bounce map,
with an effective Planck's constant $\heff= k^{-1}$ \cite{tualle}.

This observation leads to consider canonical maps $\kappa$ on simple 2-dimensional
phase spaces, and quantize them into unitary
propagators ({\it quantum maps}) $U_N(\kappa)$ of finite dimension $N\sim\heff^{-1}$ \cite{qmaps}.
A Gaussian wavepacket $|q,p\ra$ localized at the phase space point $(q,p)$
is first transformed unitarily into a deformed wavepacket $U_N\,|q,p\ra$, localized near $\kappa(q,p)$. 
To induce some damping, we then 
multiply this state by a factor $a(\kappa(q,p))$, which can be implemented by applying to
$U_N\,|q,p\ra$ the operator $\hat a$ quantizing the damping factor. The latter is
generally
complex-valued, we will assume that it satisfies the following bounds:
\be\label{e:a(z)}
\forall q,p,\quad 0<a_{\min}\leq |a(q,p)|\leq a_{\max}=1\,.
\ee
These two steps lead to the definition of the {\it damped quantum map}
$$
M_N=\Mak=\hat a\circ U_N(\ka)\,.
$$
The classical limit of the dynamics generated by $M_N$ acts on 
``weighted point particles'': a point at position $(q,p)$
is moved to $\kappa(q,p)$ and its weight is reduced by a factor $|a(\kappa(q,p))|^2$. This
is the discrete-time version of a weighted ray dynamics.
Compared with cavity systems, this model has two main advantages: 
one can easily engineer a map $\kappa$ with specific dynamical properties;
the spectrum of $\Mak$ is easier to study both numerically and analytically.

The spectrum
$\{\lambda_j^{(N)}\}_{1\leq j\leq N}$ of $\Mak$ is the main object of our study
(eigenvalues are ordered by decreasing moduli). 
To compare it with the resonance spectrum of a damped cavity,
one should extract from the latter an interval $\{|\omega_n-k|\leq\pi\}$ around 
the frequency $k\sim  N$.
The distribution 
of the decay rates $\{\Gamma_n\, :\,  |\omega_n - k|\leq\pi\}$ should parallel that of
the decay rates $\{ \gamma_j^{(N)}= -2\ln | \lambda_j^{(N)}|\}_{1\leq j\leq N}$.

\medskip

A similar model had been introduced in \cite{SchTw,NZ} to mimick 
``fully open'' cavities: 
the damping factor $a(z)$ was then vanishing inside the opening.
Such systems were characterized by a {\it fractal Weyl law} \cite{SZ1,SchTw,NZ}:
the number of resonances in a strip $\{|\omega_n|\leq k,\ \Gamma_n\leq \Gamma\}$ grew like 
$k^{1+\delta}$, where $\delta<1$ was given by the fractal dimension of the trapped set.
On the opposite, the bounds \eqref{e:a(z)} imply that for $N$ large enough, 
$M_N$ is invertible, and its $N$ eigenvalues are contained inside the
annulus $\{a_{\min}\leq |\lambda_j^{(N)}|\leq 1\}$.
Transposed to the case of an absorbing
cavity, it implies that all high-frequency resonances
are contained in a fixed strip $\{\Gamma_n\leq\Gamma_{\max}\}$, and the number of modes 
$\mathcal{N}\{k_n\;:\ |\omega_n|\leq k \}$ asymptotically grows like $C\,k^2$, thus satisfying
a {\it standard} Weyl law \cite{Sj}.
The situation is more complicated for dielectric cavities. Explicit solutions
in the case of the circular cavity \cite{Duber} suggest that 
resonances split between two well-separated groups:
``inner'' resonances contained in a strip $\{\Gamma_n\leq \Gamma\}$, and ``outer''
resonances $\Gamma_n\sim \omega_n^{1/3}$ associated with modes localized
outside the cavity. 
Since our damped quantum map only acts on states localized ``inside the torus'', we 
believe that the above Weyl asymptotics correctly counts
the inner resonances of dielectric cavities
(the fractal Weyl law recently observed in \cite{WM}
is probably a finite-frequency artifact).

\medskip

To obtain a more precise description, one needs to iterate the dynamics, that is 
study the time-$n$ evolution $M_N^{n}$. 
Applying the quantum-classical correspondence (``Egorov's theorem''), we find that
\be\label{e:iterate}
\big(\Mak^{n\dagger}\Mak^{n}\big)^{1/2n} \approx 
\hat a_n,
\ee
where the function $a_n= \big(\prod_{i=1}^n |a\circ\kappa^i|\big)^{1/n}$ is the average damping over
trajectory stretches of length $n$. The approximation is valid in the semiclassical limit $N\to\infty$.

Much can be drawn from the knowledge of the functions $-2\ln a_n$ in the long-time limit $n\gg 1$. 
Their {\it ranges} consist in intervals $I_n(a)\subset I_{n-1}(a)$, which converge to
a limit interval $I_\infty(a)$ when $n\to\infty$. The above identity implies that the quantum
decay rates $\gamma^{(N)}_j$ must be contained in $I_\infty(a)$ for large enough $N$ \cite{Sj}. 
Numerical \cite{KNSch} and analytical \cite{AL} studies
indicate that the ``quantum ranges'' $J_N(a)=[\gamma_1^{(N)},\gamma_N^{(N)}]$ generally remain
strictly inside $I_\infty(a)$, in particular stay at finite
distance from zero.
Adapting methods used to study scattering systems \cite{Gaspard-Rice,NZ2}, 
one finds that high-frequency decay rates should be larger than
$\gamma_{gap} \defeq -2\cP_\kappa(\ln|a|-\lambda^u/2)$, where $\cP_\kappa(\cdot)$
is the {\it topological pressure} associated with the map $\kappa$ and the 
observable $(\ln|a|-\lambda^u/2)$ \cite{Walters},
$\lambda^u(z)$ being the expansion rate of $\kappa$ along the
unstable direction. 
The lower bound $\gamma_{gap}$ may be trivial (negative) when $|a(z)|$ varies little across the phase space 
(see the last line in Table~\ref{t:rates}).

Since $\kappa$ is chaotic, the {\it value distribution} of 
$-2\ln a_n(z)$ becomes peaked around its average $ \gamma_{typ}= -2\int\ln |a(z)|\,dz$ when
$n\to\infty$.
The central limit theorem \cite{CLT} shows that this distribution is asymptotically
a Gaussian of width $\frac{\sigma(a)}{\sqrt{n}}$ around $\gamma_{typ}$.
This distribution is semiclassically connected with the
spectral density of $\hat a_n$: denoting $\{s_i^{(N,n)}\}_{1\leq i\leq N}$ its eigenvalues and $\mathcal V$ the 
volume on the torus, we have the Weyl law
\begin{equation}
\mathcal{N} \{ s_i^{(N,n)} \leq s\} \approx N\; \mathcal V\left( a_n^{-1}([a_{\min},s])\right)
\end{equation}
From \eqref{e:iterate}, the LHS approximately counts the
{\it singular values} of the matrix $M_N^n$.
Using the Weyl inequalities \cite{Sing-values}, we obtain that
most of the decay rates $\{\gamma_j^{(N)}\}_{1\leq j\leq N}$ 
satisfy $\gamma_j^{(N)}\geq \gamma_{typ}-\epsilon$.
 
Applying the same argument to the inverse quantum map 
$M_N^{-1}\approx M_N(a^{-1}\circ\kappa,\kappa^{-1})$, we 
eventually find that in the semiclassical limit, 
most decay rates cluster around $\gamma_{typ}$, which we thus
call the {\it typical decay rate}. More precisely, the fraction of the decay rates 
$\{\gamma_j^{(N)}\}$ which are not in the interval 
$[\gamma_{typ}-\eps,\gamma_{typ}+\eps]$ goes to zero when $N\to\infty$.

\begin{figure}[ht]
\includegraphics[width=0.48\columnwidth]{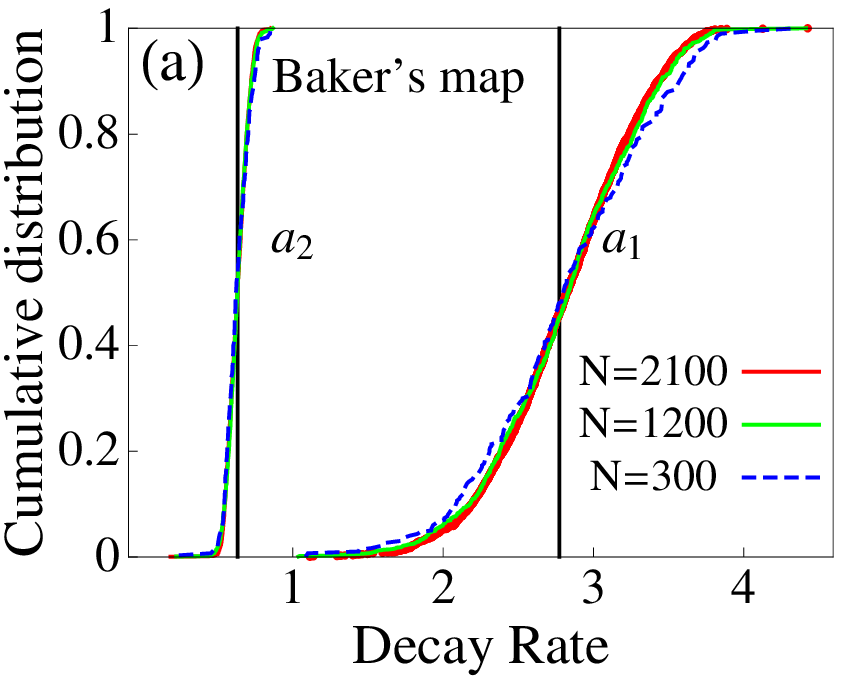}
\includegraphics[width=0.48\columnwidth]{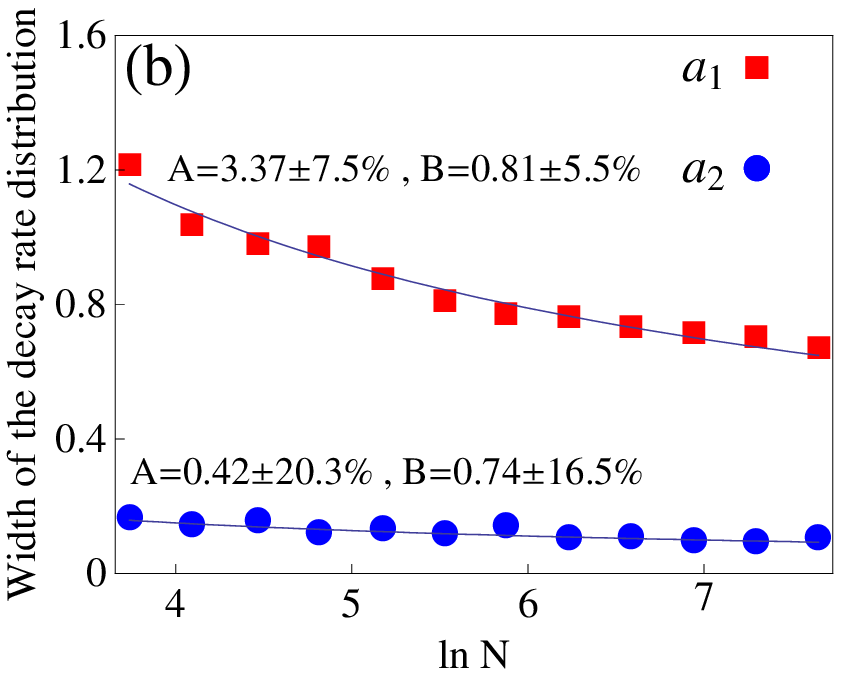}
\caption{(Color online) (a) Cumulative decay rate 
distribution for $M_N(\ka,a_i)$.
The vertical bars indicate the rates $\gamma_{typ}$. 
(b) Widths of the decay rate distributions, together with
the best 2-parameter fits  $A\,(\ln N)^{-B}$ and the asymptotic standard errors. The power-law 
fits  $A'\, N^{-B'}$ yield 
$A'=1.35\pm5\%\,,B'=0.09\pm8\%$ for $M_N(a_1)$ and $A'=0.2\pm18\%\, , B'=0.10\pm26.5\%$ for $M_N(a_2)$.
\label{f:clust}}
\end{figure}

By pushing the quantum-classical correspondence up to its limit, namely 
the {\em Ehrenfest time} $n\sim C\ln N$, we find that the {\it width} of the
decay rate distribution 
is at most of order $(\ln N)^{-1/2}$ (a rigorous proof 
will be given in \cite{Schenck}).
Our numerics (see Fig.\ref{f:clust}) are compatible with this upper bound. Such a slow
decay could explain why this concentration has not been detected in previous studies.
For a solvable toy model presented at the end of this note, the distribution
will be shown to be indeed a Gaussian of width 
$\sim C(\ln N)^{-1/2}$.

\begin{figure}
\includegraphics[width=0.49\columnwidth]{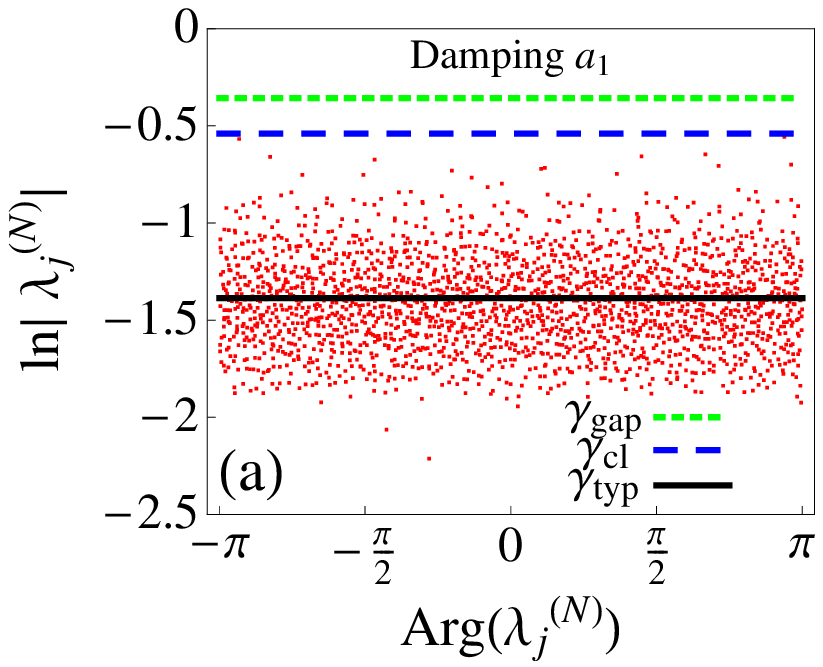}
\includegraphics[width=0.49\columnwidth]{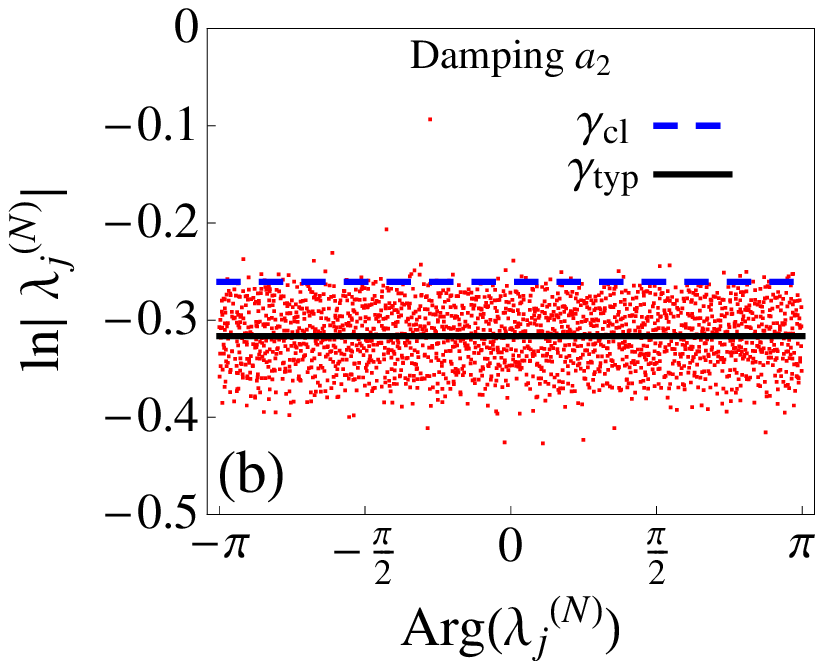}
\caption{\label{f:spectres} (Color online) Spectra of $M_N(\kappa,a_i)$ for $N=2100$ and damping factors
$a_1$ (a) and $a_2$ (b). We
plot $\{i\ln(\lambda_j^{(N)})\}$ to mimick the spectrum of a damped cavity near $k\sim N$ 
(the vertical coordinates correspond
to $-\gamma_j^{(N)}/2$). The horizontal
lines indicate $-\gamma_*/2$ for the theoretical rates given in table~\ref{t:rates}.}
\end{figure}

Let us compare the quantum decay rates with
the ``classical decay rate'' $\gamma_{cl}$ of the corresponding weighted dynamics. The latter,
introduced in \cite{Lee04} in the framework of dielectric microcavities, 
is obtained by evolving an initial smooth distribution of points through the
weighted dynamics: for large times $n$, the total weight of the distribution decays like $W\,e^{-n\gamma_{cl}}$.
As in the case of fully open systems \cite{Gaspard-Rice}, 
$\gamma_{cl}$ can be expressed as the topological pressure $\gamma_{cl}=-\cP_\kappa(2\ln|a|-\lambda^u)$.
Convexity properties of the pressure allow to compare this classical
decay rate with the two rates obtained above: $\gamma_{gap}\leq \gamma_{cl} \leq \gamma_{typ}$,
and the inequalities are generally strict (see table~\ref{t:rates}). The quantum ranges $J_N(a)$ 
may or may not contain the classical rate $\gamma_{cl}$ (see Fig.~\ref{f:spectres}).

\medskip

The map we consider in our numerics is the 3-baker's map, which acts
canonically on the 2-dimensional torus $\{(q,p)\in [0,1)^2\}$.  It is given by 
$\ka(q,p)=(3q\bmod 1, \frac{p+[3q]}{3})$, and generates a strongly chaotic dynamics. 
This map is quantized as in \cite{Sarac}, into a sequence of unitary matrices
$U_N=G_N^{-1}\begin{pmatrix}G_{N/3}&&\\&G_{N/3}&\\&&G_{N/3}\end{pmatrix}$, where
$(G_M)_{jk}=\frac{1}{\sqrt{M}}\,\exp\big(-\frac{2i\pi}{M} (j+\frac12)(k+\frac12)\big)$ is the ``symmetrized'' 
discrete Fourier transform.
We choose damping factors of the form $a(q)$, so their quantizations
$\hat a$ are simply diagonal matrices with entries $a((j+1/2)/N)$. 
The factor $a_1(q)$ has a plateau 
$a_1(q) \equiv 1$ for $q\in [1/3, 2/3]$, another one $a_1(q) \equiv 0.1$ 
for $q\in [0, 1/6]\cup [ 5/6, 1]$, and varies smoothly inbetween. It approximates
the piecewise constant function $\tilde a_1(q)$ which takes values $0.1$, $1$, and $0.1$ respectively
on the intervals $[0,1/3)$, $[1/3,2/3)$ and $[2/3,1)$.
Our second choice is the ``smoother'' function 
$a_2(q)=1-\sin(2\pi q)^2/2$. Since we use a single map $\kappa$, the damped quantum maps will
be abbreviated by $M_N(a_i)$.

We first notice that all these factors reach their extremal values $a_{\min},\,a_{\max}$ on the 
fixed points $(0,0)$ and $(1/2,1/2)$ of $\ka$.
As a result, for each of them the asymptotic range $I_\infty(a)$ is equal to $[a_{\min},a_{\max}]$.
The theoretical rates $\gamma_{gap}$, $\gamma_{cl}$ and $\gamma_{typ}$ for these three factors are 
given in Table~\ref{t:rates}.
\begin{table}
\begin{tabular} {|c||c|c|c|}
\hline
  &$\gamma_{gap}$&$\gamma_{cl}$&$\gamma_{typ}$\\
\hline
$a_1$&      0.715      &   1.079        &2.774      \\
\hline
$\tilde a_1$& 0.734   &   1.079        &3.070      \\
\hline
$a_2$&     -0.523      &   0.521        &0.633  \\
\hline
\end{tabular}
\caption{Values of the theoretical rates for $\kappa$ and the various damping functions we use.
\label{t:rates}}
\end{table}
In Fig.\ref{f:spectres} we plot the spectra of $M_N(a_i)$ for $N=2100$ (the theoretical bound 
$\gamma_{gap}$ for $a_2$ is negative, hence irrelevant). We check that
all quantum rates are larger than $\gamma_{gap}$. In the case of $M_N(a_1)$, all quantum rates are
also larger than $\gamma_{cl}$, while $M_N(a_2)$ admits a few smaller decay rates.

The clustering of decay rates around $\gamma_{typ}$ is already perceptible in Fig.\ref{f:spectres}. To make
it more quantitative, 
in Fig.\ref{f:clust} (a) we plot the cumulative distributions of decay rates.
At first glance, the widths of the distributions
around $\gamma_{typ}$ seem to depend little on $N$.
Enlarging the set of data, we plot these widths on Fig.\ref{f:clust} (b).
They indeed decay with $N$. The 2-parameter power-law fits $A'\,N^{-B'}$ lead to small exponents $B'$,
which seem to favor the logarithmic fits $A\,(\ln N)^{-B}$: the latter
decay slightly faster than the theoretical upper bound $(\ln N)^{-1/2}$.

It is possible to construct a {\em solvable} quantization of the baker's map by taking
the quantum parameter $N=3^k$, $k\in\IN$, and replacing the
discrete Fourier transform $G_N$ by the Walsh transform \cite{NZ}.
If we then select a
damping factor which, like $\tilde a_1$, takes constant values $a^j$ on the intervals 
$q\in[j/3,(j+1)/3)$, 
the quantum model remains solvable. The spectrum of $M_N$ relies on
the eigenvalues $\{\lambda_i\}$  of the $3\times 3$ matrix ${\rm diag}(a^j) G_3^{-1}$.
Taking $\gamma_i=-2\ln|\lambda_i|$, the $N$ quantum decay rates can be indexed by
the sequences $\bet=\eta_1\eta_2\ldots\eta_k$, with $\eta_i\in\{1,2,3\}$: they are given by
$\gamma_{\bet}=\frac1{k} \sum_{m=1}^k \gamma_{\eta_m}$.
For instance, in the case of the damping function $\tilde a_1$, the rates $\gamma_i$ take the values
$(0.803,\,3.801,\,4.605)$. 
From this explicit expression, one easily draws that, in the limit $k\to\infty$, 
the distribution of the $\{\gamma_{\bet}\}$ converges to a Gaussian
of average $\gamma_{typ}=(\sum_{i=1}^3 \gamma_i)/3$ and variance 
$\frac{1}{3k}\sum_{i=1}^3 (\gamma_i-\gamma_{typ})^2=C (\ln N)^{-1}$.

\medskip

To summarize, we have studied the spectra and eigenmodes of damped quantum chaotic maps, a 
toy model for various types of partially open quantized chaotic cavities, in a r\'egime
where the damping is both macroscopic and strongly nonperturbative. We have shown that the
quantum decay rates remain inside a fixed interval, and that most of them cluster around 
the mean damping rate $\gamma_{typ}$. 
These statistical properties seem to differ from those
of non-hermitian random matrices used to represent such open systems. 

{\it Acknowledgements:} This work was partially supported by the grant ANR-05-JCJC-0107-01. 
We have benefitted from discussions with
F.Faure and Y.Fyodorov, and thank E.Bogomolny and R.Dubertrand for sharing their
results prior to publication.



\begin{thebibliography}{99}
\bibitem{Sing-values} Assume the singular values $\{s_i\}$ and eigenvalues 
$\{\lambda_i\}$ of a $N\times N$ matrix
are ordered by decreasing moduli. Then, for any $t>0$ and any $j\leq N$, one has 
$\sum_{i=1}^j|\lambda_i|^t\leq \sum_{i=1}^j s_i^t$.
\end{thebibliography}
\end{document}